# Programmable Quantum Mode Switches via Plasmonic Toroidal Nanoantennae

Arda Gulucu[1,2] and Emre Ozan Polat[1,2,*]

[1]*Department of Physics, Bilkent University, 06800, Ankara, Turkey*

[2]*UNAM - National Nanotechnology Research Center and Institute of Materials Science and Nanotechnology, Bilkent University, 06800, Ankara, Turkey*

**Corresponding author: emre.polat@bilkent.edu.tr*

## Abstract

The ability to switch and program the spectral response of quantum modes via deterministically located plasmonic nanoantennae presents opportunities for wide spectrum of applications from biosensors to quantum computing. Due to its topology, toroidal nanoantenna (TNA) focuses immense amount of three-dimensional (3D) local electric field by toroidal moment while allowing pre and post positioning around quantum emitters (QEs). Here, within local-response finite difference time domain (FDTD) simulations, we demonstrate high-contrast spectral switching of the radiative decay channel of a dipolar QE coupled to a TNA by introducing effective Lorentzian quantum objects (QOs). At optimized TNA geometries, Fano interference between the broadband plasmonic continuum and narrow quantum transitions of QOs suppresses both radiative and non-radiative decay channels near 850 nm, yielding an observable full switching that traps energy within the hybrid mode instead of re-emitting it. To show the promises of the concept, we further demonstrate systems with multiple QOs where spectral degeneracy enhances the transparency bandwidth, while detuning generates distinct minima, enabling individually addressable spectral responses. These results establish plasmonic TNAs as promising architectures for spectral detection and individual mode switching of single- or multi-QO configurations and empowers the user for the implementation of photonic processing of continuous photon sources.

**Keywords:** Toroidal nanoantenna, Fano interference, Purcell enhancement, Quantum emitter, Plasmonic mode switching

## Introduction

Plasmonics exploits the collective oscillations of free electrons at metal-dielectric interfaces to concentrate electromagnetic energy into nanoscale volumes known as "hotspots" [1,2]. Localized surface plasmon resonances (LSPRs) in metallic nanoparticles generate intense electric fields near their surfaces, enabling surface-enhanced Raman scattering, fluorescence modulation, and nanoscale sensing [3–6]. Spherical and rod-shaped nanoparticles have been extensively studied in the context of field localization [7–9]. Unlike commonly used spheres, rods or rings, TNA supports closed poloidal displacement currents that realize a toroidal (anapole) moment, squeezing fields into the inner rim with small mode volume [10,11]. Offering unique confinement properties, the topology of a metallic TNA produces a 3D hotspot volume with simultaneous radial and axial localization, supported by circulating plasmon modes [12–16]. This configuration provides versatile tunability of resonance wavelength through its aspect ratio while delivering a high local density of optical states (Purcell enhancement [17]) and multipolar mode mixing [18]. These features make toroidal plasmonic resonators a strong contender for engineering strong light-matter interactions at deterministic emitter locations.

When a narrow discrete resonance interacts coherently with a broad plasmonic continuum, their interference gives rise to an asymmetric Fano lineshape, manifested as a sharp dip or transparency window within the otherwise broad scattering spectrum [19,20]. The discrete state can originate from QOs such as molecules (or molecule arrangements), quantum dots (QDs), or excitons possessing intrinsically narrow linewidths [21–25]. These hybrid plasmon-QO Fano resonances produce spectrally sharp and tunable features with enhanced local fields and strong dispersion enabling promising results in high-sensitivity sensing, optical switching, low-threshold lasing, and nonlinear photonic functionalities [26,27].

Here, we introduce an active mode-switching mechanism for a quantum emitter (QE) coupled to a toroidal nanoantenna (TNA), enabling discrete quantum modes to be selectively controlled across a strongly enhanced photoluminescence (PL) spectrum. Although TNAs have attracted significant attention, prior work has largely remained within the framework of passive resonance engineering mainly focusing on Purcell enhancement, radiation efficiency, and emission directivity [12–16]. Similarly, finite-element studies comparing single-torus and dimer geometries quantified linewidth, near-field enhancement, and Purcell factors [28], but did not address whether a radiatively optimized TNA could be transformed into an actively programmable interference platform.

On the other hand, Fano resonance-based studies demonstrated the interference of discrete modes and the resulting modulation at specific wavelength regimes, but they did not achieve complete switching of the emission channel to be used as a quantum mode switch [29–31]. Here, we show that a single Lorentzian QO, tuned into resonance with the QE emission, can completely suppress the radiative decay channel that is otherwise enhanced by the factor of 2840 in the vicinity of the TNA. The key novelty is that the Lorentzian QO does not need to act as a bright emitter itself; instead, it switches the TNA-enhanced decay pathways by sharply perturbing the local electromagnetic response through the frequency-dependent dielectric response, ε(ω). This makes the concept broadly applicable as a general and programmable platform for emission control over plasmonic continuum, with potential relevance to defect centers [32,33], various types of molecular excitonic assemblies [34], QDs [35], and any QO that possess intrinsically narrow linewidths [19,20].

We first report a large span of emitter-antenna distances where the QE radiative decay rate dominates over non-radiative counterpart for a specific TNA aspect ratio regime. Then, we simulate the hybrid system consisting of QOs, TNAs and QEs through FDTD simulations and report an efficient quantum mode switching mechanism relying on a Fano transparency in the enhanced spectrum, yielding a full switching of quantum modes over the plasmonic continuum.

Composed of a plasmonic nanoantenna and a deterministically located QO, the demonstrated quantum mode switches do not require complex multi-resonator unit cells or three-dimensional metamaterial architectures that are commonly employed by the previous reports [36–39]. The experimental implementation of the approach is viable through top-down [40] and bottom-up [41] nanofabrication methods provided that the toroidal structures can be fabricated around QEs, and the QOs can be imposed later. Alternatively, the TNA can be located around a defect center [42], ensuring deterministic alignment [43,44].

To show the promises of the approach, the investigation is extended from single QO to multi-QO configurations demonstrating that spectral degeneracy enhances the transparency bandwidth, while detuning among multiple QO (such as the Stark effect [45]) generates multiple minima that are individually addressable over the plasmonic continuum. The reported results provide a benchmark for label-free spectral sensing in the single particle limit [46] and spectral control of individually addressable quantum modes [47]. We anticipate that the experimental demonstration of the presented system would remain niche advancements for super resolution bioimaging [48], quantum sensing [49], programmable photonic circuits [50] and future display technologies [51].

**Results**

We first consider a system composed of a dipolar QE placed at a distance $d$ from a silver TNA characterized by major radius $R$ and cross-sectional radius $\alpha$ (**Figure 1a**). **Figure 1b** presents a snapshot of the electric-field intensity distribution at $t \approx 0.14$ps, showing that the dipole field emitted by the QE becomes strongly localized around the TNA. By varying the aspect ratio $\alpha/R$, the plasmonic resonance wavelength $\lambda_{\text{res}}$ of the TNA can be systematically tuned. For a fixed total radius size of $R + \alpha = 60$ nm, increasing $\alpha/R$ from 0.1 to 1 shifts $\lambda_{\text{res}}$ from the near-infrared to the UV-visible region (**Figure 1c**), demonstrating that the same TNA platform can be adapted to a broad range of QE transition energies. Consistent with analytical expectations, decreasing the aspect ratio reduces the effective toroidal volume and produces a red shift in the resonance wavelength [12]. In contrast, as $\alpha/R$ approaches unity, the toroid increasingly resembles a metallic nanosphere of radius $R+\alpha$, causing $\lambda_{\text{res}}$ to saturate for $\alpha/R \gtrsim 0.7$ (**Figure 1c**).

**Figure 1d** shows the normalized radiative and non-radiative decay rates for an emitter-antenna separation of $d = 3$ nm and a fixed total radius $R + \alpha = 60$ nm. Although non-radiative decay is often expected to dominate at such short distances because of near-field energy transfer to the metal, we identify a counterintuitive regime in which the radiative decay channel exceeds the non-radiative one. In particular, the radiative rate reaches its maximum near $\alpha/R = 0.2$, and remains larger than the non-radiative contribution over the range $0.2 \lesssim \alpha/R \lesssim 0.6$, as highlighted in the inset of **Figure 1d**.

This intermediate regime can be understood analytically through the Purcell factor,

$$F_P = \frac{\gamma_{\text{tot}}}{\gamma_0} = \frac{3}{4\pi^2}\left(\frac{\lambda}{n}\right)^3 \frac{Q}{V_{\text{eff}}}, \qquad (1)$$

where $\gamma_{\text{tot}} = \gamma_r + \gamma_{nr}$ is the total decay rate of the emitter in the presence of the nanoantenna, $\gamma_0$is the free-space decay rate, $\lambda$is the wavelength, $n$is the refractive index of the surrounding medium, $Q$is the quality factor, and $V_{\text{eff}}$ is the effective mode volume. Decreasing the aspect ratio produces two concurrent effects: it redshifts the resonance to longer wavelengths and reduces the effective mode volume. Together, these changes increase the LDOS and thus enhance the Purcell factor, while the accompanying reduction in $Q$ remains comparatively minor.

Within the optimal range, $0.2 \leq \alpha/R \leq 0.6$, the LDOS enhancement couples predominantly to radiative channels, so the plasmonic TNA effectively operates as a cavity despite its modest quality factor. For smaller aspect ratios, $\alpha/R < 0.2$, the system enters a Rayleigh-like absorption regime in which losses dominate. For larger aspect ratios, $\alpha/R > 0.6$, the response gradually transitions toward that of a silver nanosphere, where Ohmic dissipation becomes dominant. The sharp increase in non-radiative decay beyond $\alpha/R \approx 0.6$ is therefore best interpreted as a modal crossover rather than a smooth continuation of the same toroidal mode.

As shown in **Supplementary Figures S1-S3**, although a toroid-like resonance persists for $\alpha/R = 0.7$-1.0, the dominant plasmonic response shifts to a sphere-like mode near 342 nm, which is strongly non-radiative. As the inner opening of the toroid closes, the structure increasingly resembles a compact nanosphere, weakening the radiatively efficient toroidal resonance and favoring Ohmic loss. Hence, the abrupt rise in $\gamma_{nr}$ does not arise from a gradual modification of the original toroidal mode, but from the emergence of a loss-dominated sphere-like plasmon branch. The corresponding values of $\lambda_{res}$ and the normalized decay rates for different aspect ratios are summarized in **Table 1**.

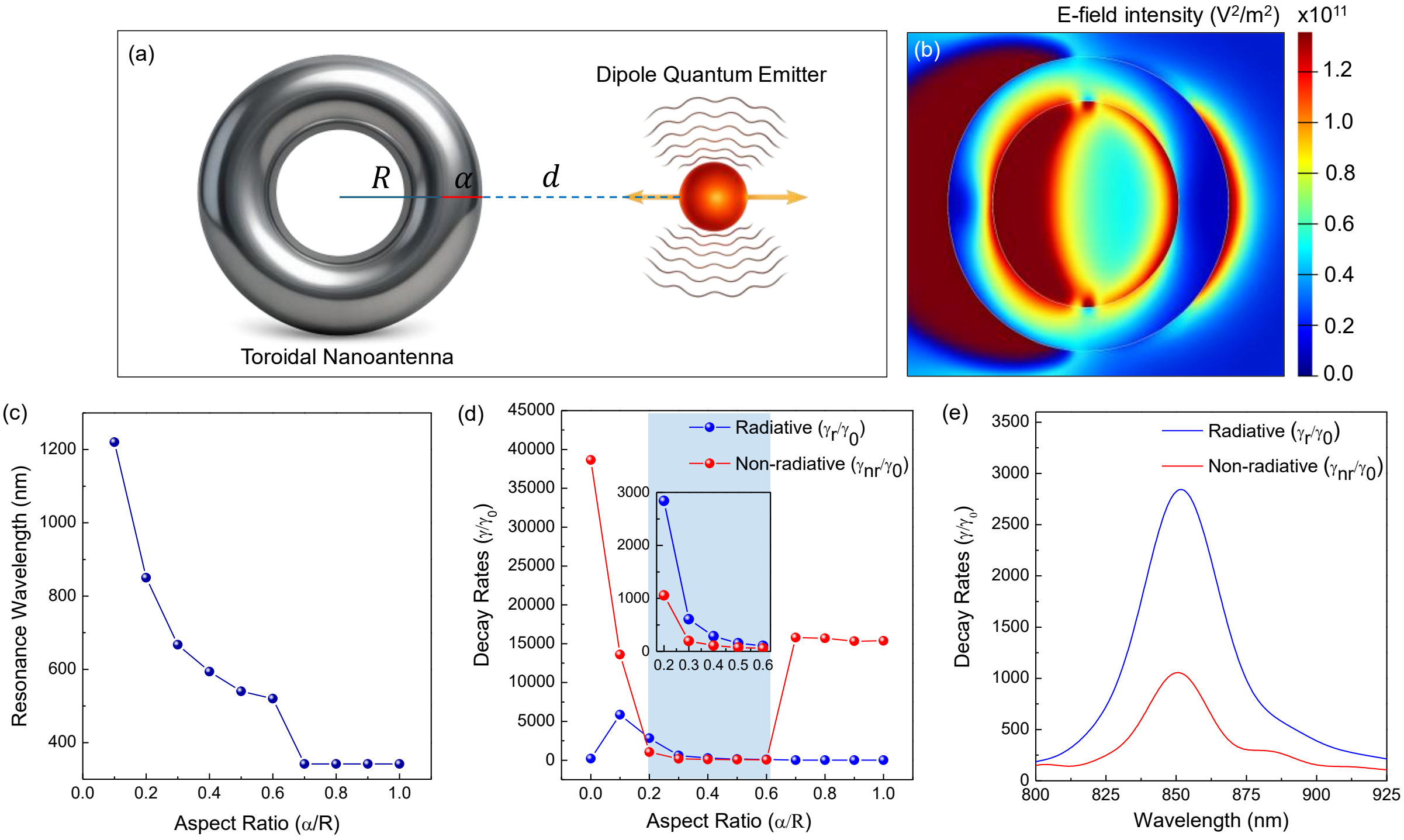

**Figure 1**. **Geometry-controlled radiative enhancement and spectral tunability in a quantum emitter-toroidal nanoantenna system**. **(a)** Schematic illustration of the quantum emitter-toroidal nanoantenna (QE-TNA) system for plasmonic enhancement. A dipole QE (red) located at distance d away from a silver TNA, with a major radius R and cross-section radius α. **(b)** Snapshot of local electric field intensity ($|E_{loc}|^2$) at t ≈ 0.14 ps for the presented QE-TNA system. Color bar shows the E-field intensity scale. **(c)** Resonance wavelengths for increasing aspect ratios (α/R). By changing the ratio between 0.2 and 1, one can yield a broadband range of resonance wavelengths from NIR to UV-Vis **(d)** Normalized decay rate ($\gamma/\gamma_0$) values at resonance wavelength (850 nm) with respect to TNA aspect ratio (for d = 3 nm). Inset shows the intermediate aspect ratio regime where the radiative decay rates ($\gamma_r/\gamma_0$) overcome the non-radiative ($\gamma_{nr}/\gamma_0$) counterparts. **(e)** Decay rate spectra of QE-TNA coupling. For the optimal condition of α/R = 0.2 the enhancement of LDOS boosts the radiative decay rate over the non-radiative counterpart around the resonance wavelength of $\lambda_{res}$ = 850 nm.

**Table 1. Complete parameter set for the dipole QE-TNA coupling as a function of toroidal aspect ratio**. Listed are the plasmon resonance wavelength ($\lambda_{\text{res}}$), normalized radiative decay rate ($\gamma_r/\gamma_0$), and normalized non-radiative decay rate ($\gamma_{nr}/\gamma_0$) for different values of the aspect ratio ($\alpha/R$). The results identify an optimal intermediate geometry for radiative enhancement and a high-aspect-ratio regime characterized by increased non-radiative loss and resonance saturation.

| *Aspect Ratio* **(α/R)** | *Plasmon Resonance Wavelength* $\boldsymbol{\lambda_{res}}$ **(nm)** | *Normalized Radiative Decay Rate* $\boldsymbol{\gamma_r/\gamma_0\ (\times 10^3)}$ | *Normalized Non-radiative Decay Rate* $\boldsymbol{\gamma_{nr}/\gamma_0 (\times 10^3)}$ |
|---|---|---|---|
| 0.1 | 1220 | 5.870 | 13.60 |
| 0.2 | 850 | 2.840 | 1.056 |
| 0.3 | 667 | 0.608 | 0.195 |
| 0.4 | 600 | 0.284 | 0.109 |
| 0.5 | 542 | 0.155 | 0.073 |
| 0.6 | 514 | 0.104 | 0.056 |
| 0.7 | 342 | 0.020 | 15.80 |
| 0.8 | 342 | 0.015 | 15.70 |
| 0.9 | 342 | 0.017 | 15.33 |
| 1.0 | 342 | 0.013 | 15.38 |

We therefore identify $\alpha/R = 0.2$ as the optimal geometry for maximizing radiative efficiency, yielding a strongly enhanced radiative decay rate of $\gamma_r/\gamma_0 = 2840$ while keeping non-radiative losses comparatively lower at $\gamma_{nr}/\gamma_0 = 1056$, both centered around $\lambda_{\text{res}} = 850$nm (**Figure 1e**). Although an individual quantum emitter typically possesses a narrow Lorentzian emission linewidth, the broad spectral feature observed between 800 and 900 nm in **Figure 1e** does not represent the intrinsic emission profile of the emitter. Rather, it reflects the Purcell enhancement spectrum imposed by the plasmonic environment. In other words, the TNA provides a broadband plasmonic continuum that reshapes the LDOS experienced by the emitter, and the resulting radiative and non-radiative decay spectra inherit this broadband character. This result shows that a radiatively optimized TNA can be converted from a broadband Purcell enhancer into an actively switchable Fano platform by coupling it to a single narrow-linewidth molecular-like QO resonance.

Next, we place a single effective Lorentzian QO, modeled by a sharp Lorentzian dielectric response, at the center of the TNA (**Figure 2a**). In the present FDTD model, this QO is represented as a spherical dispersive dielectric nanoinclusion rather than an atomistically resolved single molecule. This effective description is used to capture a narrow resonant polarizability that can interfere with the broadband plasmonic continuum of the TNA. When coupled to the broadband plasmonic continuum of the QE-TNA system, the QO-induced polarization introduces a narrow spectral channel that interferes with the broad antenna response, forming a hybrid system that supports Fano-type interference between discrete and continuum states. This interference modifies the optical response of the system and generates the characteristic asymmetric Fano line shape. **Figures 2b** and **2c** show the normalized radiative and non-radiative decay spectra, respectively. In the bare QE-TNA system, both $\gamma_r$ and $\gamma_{nr}$ vary smoothly, consistent with conventional broadband plasmonic dissipation. However, when a QO of 20 nm radius is introduced at the center, a sharp suppression emerges near $\lambda = 850$ nm, marking the onset of the Fano transparency window. The simultaneous reduction of both decay channels indicates that destructive interference at this wavelength suppresses the LDOS and inhibits energy transfer into both radiative and dissipative pathways.

Importantly, the switching behavior is far more pronounced in the radiative channel than in the non-radiative one. The radiative decay is completely switched off, collapsing from $2840\gamma_0$ to nearly zero, whereas the non-radiative decay is reduced from $1056\gamma_0$ to $249\gamma_0$. This asymmetry reflects the fundamentally irreversible nature of material losses, which remain limited by electron-phonon and phonon-phonon dissipation even when radiative emission is fully suppressed. **Figure 2d** further confirms this interpretation through near-field intensity maps taken at resonance and off-resonance wavelengths around the transparency window. Strongly localized fields are observed around the TNA surface away from the transparency condition, whereas at $\lambda = 850$ nm the field intensity collapses dramatically due to destructive interference between the QO-induced polarization and the plasmonic continuum. This collapse directly visualizes the optical switching action of the Fano resonance and the associated suppression of the LDOS.

The overall Purcell response is summarized in **Figure 2e**, which compares the Purcell factor spectra of the bare QE–TNA system (orange curve) and the hybrid QE-TNA-QO system (cyan curve). The introduction of the narrow QO resonance transforms the smooth broadband continuum into an asymmetric spectral response with a pronounced transparency dip. This minimum corresponds to the destructive interference point, where the real part of the QO self-energy also induces a slight resonance shift from 348 THz to approximately 353 THz, consistent with the expected signature of Fano interference between discrete and continuum states [52]. Unlike previous Fano-based plasmonic implementations that often rely on complex cavity constraints or multi-resonator architectures, the present system achieves strong enhancement and near-complete radiative suppression in a compact toroidal platform. This simplicity is particularly attractive for scalable multipixel arrangements in which individual elements may be independently addressed [53,54].

To further clarify the physical behavior of the system and to provide a practically useful framework for experimental implementation, we next investigate the dependence on emitter-antenna separation $d$. The normalized decay rates are computed as a function of distance and fitted with an exponential form to capture the resonance dynamics of the hybrid structure. **Figure 3a** shows the normalized radiative and non-radiative decay rates at 850 nm as a function of emitter-antenna separation in the absence of the central QO. The extracted peak values of $\gamma_r$ and $\gamma_{nr}$ are fitted using an exponential distance-dependent function of the form $y(d) = e^{a+bd+cd^2}$, where $a$, $b$, and $c$ are fitting constants, consistent with previous numerical and analytical approaches (See **Methodology**) [6,55].

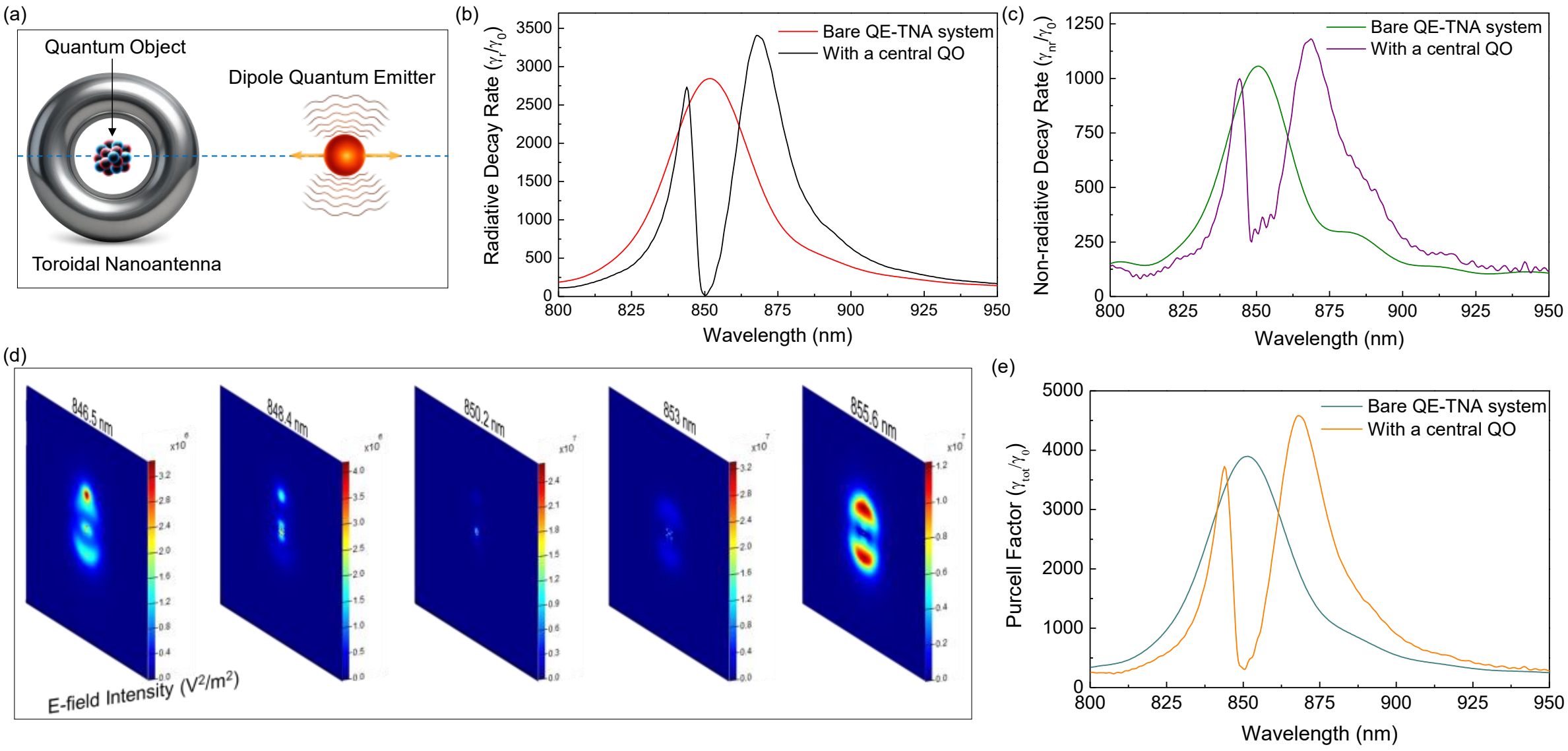

**Figure 2. Active suppression of plasmon-enhanced decay through Fano interference in the QE-TNA-QO hybrid system. (a)** Schematic of the hybrid system depicting the QE-TNA coupling to a central effective Lorentzian QO. The dispersive QO introduces a phase-shifted polarization that interferes with the broadband plasmonic continuum of the TNA, producing a Fano-type hybrid mode. **(b)** Normalized radiative decay rate $(\gamma_r/\gamma_0)$ of the system with and without QO. **(c)** Normalized non-radiative decay rate of the system with and without QO. The introduction of the QO leads to a sharp suppression in both radiative and non-radiative channels around $\lambda = 850$ nm, indicating destructive interference and energy trapping within the hybrid mode. **(d)** Electric field intensity maps at resonant and off-resonant wavelengths around the transparency region. The collapse of near-field enhancement at the resonance and the enhancement at the off-resonance wavelengths around the transparency window are clearly visible. **(e)** Purcell factor spectra of the system. The emergence of a narrow transparency dip in the Purcell factor confirms the Fano interference between the discrete QO resonance and the plasmonic continuum.

For large emitter-antenna separations, $d \gg 50$ nm, both the radiative and non-radiative decay rates approach unity, indicating recovery of the free-space emission regime. As the emitter moves closer to the toroidal nanoantenna, the non-radiative decay increases gradually, whereas the radiative decay rises sharply at very small gaps, which is unusual for metal-based plasmonic systems. Notably, the radiative channel remains dominant over the entire distance range considered. This behavior originates from the toroidal geometry and the optimized aspect ratio of the antenna, which together promote radiative LDOS enhancement while suppressing the loss-dominated near-field transfer pathways that typically emerge in conventional plasmonic structures. Close emitter-metal coupling usually involves a trade-off between plasmonic enhancement and non-radiative quenching. Previous theoretical work on emitters near metallic nanoparticles showed that the short-range non-radiative decay can follow a Förster-like dependence, while experiments on single molecules and fluorophores near gold nanoparticles confirmed strong distance-dependent quenching at small separations [55–58]. Against this background, the optimized TNA geometry studied here is notable because the simulated radiative decay remains larger than the non-radiative decay over $d = 3$–$50$nm. We therefore interpret the structure as a radiatively efficient local-response geometry, without claiming that toroidal topology universally eliminates quenching relative to all conventional plasmonic antennas.

**Figure 3b** shows the radiative decay spectra for different QE-TNA separations ($d = 3, 4, 5, 10, 15, 20, 25, 30, 40,$ and $50$ nm) in the presence of the central QO. Each spectrum exhibits a characteristic Fano transparency dip near 850 nm, although the modulation depth varies with distance. As separation increases, the dipole-plasmon coupling weakens, and the destructive interference responsible for the Fano cancellation becomes progressively less pronounced. This behavior is quantified in **Figure 3c** through the transparency contrast, $\Delta I/I\%$, defined as the relative suppression of the radiative decay rate at the resonance wavelength. At $d = 3$ nm, the transparency reaches 99.9% and remains essentially unchanged up to $d = 10$ nm, thereby defining a practically useful distance window for complete switching of the radiative channel. Beyond this range, the transparency decreases only slightly, reaching 97.8% even at $d = 50$ nm. The persistence of this near-complete switching over such a wide distance range provides substantial design flexibility for experimental implementation.

Previous emitter-nanoantenna studies in the same near-infrared spectral region studied light-harvesting complexes coupled to resonant gold nanorod antennas and reported fluorescence enhancement under 850 nm excitation, with lifetime shortening attributed to both radiative-rate enhancement and additional non-radiative loss [59]. In a related single-complex experiment, antenna-enhanced LH2 emission near 870 nm showed more than 500-fold fluorescence enhancement, while simulations predicted radiative-rate enhancements on the order of $10^2$ for nanometer-scale emitter–antenna separations [60]. Compared with these conventional nanorod systems, the larger radiative enhancement predicted here should be interpreted as a local-response FDTD result of the optimized toroidal geometry, rather than as a directly measured fluorescence enhancement factor.

We next extend the analysis beyond the single-QO case, motivated by the fact that practical implementations will generally require multiple molecular-like resonances. The single central QO establishes the fundamental mechanism: Fano interference between a narrow quantum-object resonance and the broadband toroidal plasmonic continuum. Building on this basis, multiple QOs can introduce additional interfering narrow resonances and thereby enrich the spectral response. In principle, a shared plasmonic mode may also mediate emitter-emitter interactions through the electromagnetic Green's function. In the present work, however, we adopt an effective independent-response description in which each QO interacts primarily with the same toroidal plasmonic background. Accordingly, the multi-QO features discussed below are interpreted as collective spectral modulation generated by several narrow resonances interfering with a common broadband continuum, rather than as explicit plasmon-mediated entanglement. To capture this behavior, we place additional QOs within the three-dimensional hotspot region of the TNA and examine how inter-QO spacing and spectral detuning affect the resulting switching dynamics.

To extract the decay spectra and the corresponding Fano-switching behavior for single- and multiple-QO configurations, we fix the dipole QE at $d = 3$ nm and keep one QO (QO-1) at the center of the TNA. Three additional external QOs, each with a radius of 20 nm, are then positioned around the TNA with a lateral separation of 5 nm in different directions (**Figure 4a**). When all QOs are resonant at 850 nm, we find that the Fano transparency bandwidth increases systematically with the number of coupled QOs (**Figure 4b**). While a single QO yields a transparency bandwidth of 14 nm, the full width at half maximum (FWHM) broadens to 22, 27, and 34 nm for 2, 3, and 4 QOs, respectively. Importantly, this broadening occurs without a significant shift in the transparency wavelength, indicating that the resonance frequency remains essentially fixed while the spectral switching window becomes wider. This behavior arises because multiple discrete resonances reinforce destructive interference within the same plasmonic continuum. A slight increase in transparency depth is also observed with increasing QO number; however, this effect remains modest because the dipole QE-TNA system is already operated under near-optimal conditions, yielding a maximum transparency of 99.9% at $d = 3$ nm for $\alpha/R = 0.2$ (See **Figure 3c** and **Table 1**).

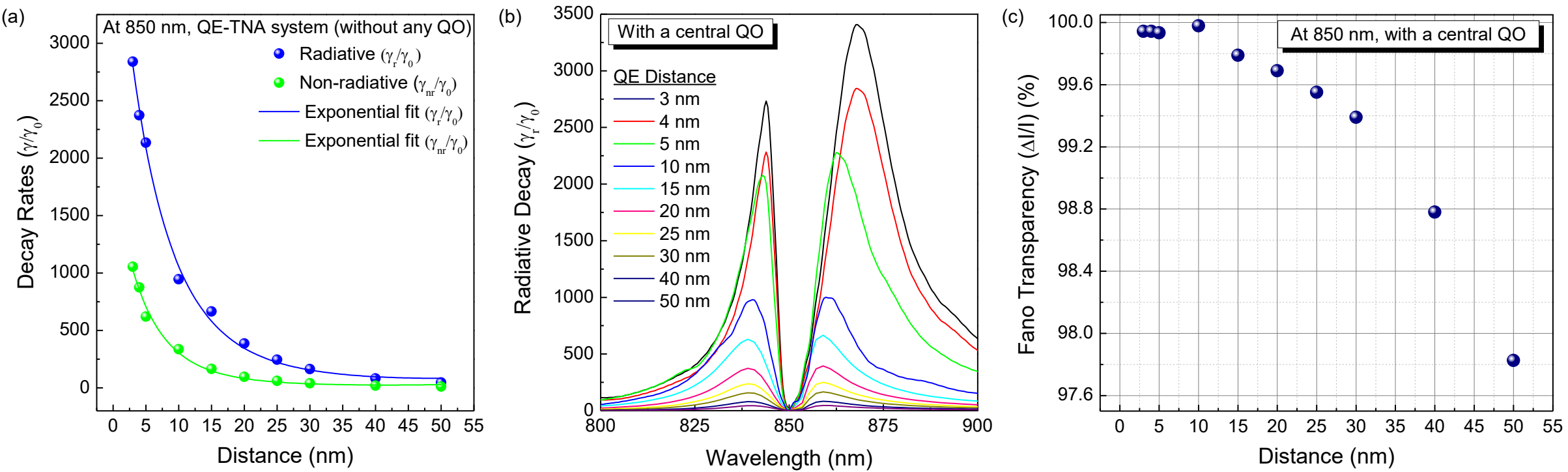

**Figure 3. Distance-tolerant Fano switching enabled by the radiatively optimized toroidal nanoantenna geometry. (a)** Normalized radiative and non-radiative decay rates at the QE emission wavelength (850 nm) as a function of emitter-antenna separation, in the absence of the central QO. Owing to the toroidal topology and optimized antenna geometry, the radiative decay remains higher than the non-radiative decay over the full distance range ($d$ = 3-50 nm), thereby preserving observable emission enhancement even at small separations. The exponential fits indicate convergence toward the free-space limit at large distances. **(b)** Radiative decay spectra for different emitter-antenna separations in the presence of the central QO. All spectra exhibit a characteristic Fano transparency dip near 850 nm, where the Lorentzian QO resonance is aligned with the QE emission. As $d$ increases, the dipole-plasmon coupling weakens and the interference contrast gradually decreases. **(c)** Fano transparency contrast, expressed as the percentage suppression of the radiative decay, $\Delta I/I\%$, as a function of emitter-antenna separation. Complete switching is maintained below d = 10 nm, while the modulation depth decreases only slightly from 99.9% to 97.8% over the $d$ = 10-50 nm range, demonstrating a robust and practically useful switching regime.

Having established the behavior of multiple resonant QOs, we next examine the effect of spectral detuning between them. In realistic systems, perfect spectral alignment among multiple quantum objects is rarely maintained, since local environmental disorder, applied electric fields, or mechanical strain can shift their transition energies. For example, the quantum-confined Stark effect (QCSE) is widely used to tune the emission or absorption wavelength of quantum dots and single molecules through static electric fields [61] . Similar spectral tuning can also be achieved intentionally by strain engineering, as demonstrated for quantum dots and layered-material emitters [62]. **Figure 4c** shows the radiative decay spectra for a double-QO configuration consisting of QO-1 and QO-2 under different detuning conditions. For simplicity, QO-1 is kept fixed at the resonance wavelength of 850 nm, while QO-2 is progressively detuned. We find that detuning one of the QOs produces a pronounced modification of the plasmonic continuum, generating a second transparency dip that shifts toward longer wavelengths as the detuning increases. For tuning values of 26.4, 32.8, and 48.6 meV, the second transparency window associated with QO-2 appears at 865, 870, and 880 nm, respectively, with bandwidths of approximately 10 nm.

To demonstrate the practical potential of this approach, we finally consider multiple detuned QOs distributed within the three-dimensional hotspot region of the TNA, as would be relevant for applications requiring the selective sensing or suppression of spectral signals from specific QOs. **Figure 4d** shows the spectral fingerprints of four QOs encoded as distinct transparency dips in the radiative decay spectrum of the hybrid system. When spectrally detuned QOs are positioned at fixed locations, each one generates its own Fano dip at its corresponding resonance wavelength, demonstrating that multiple spectral minima can coexist within the same plasmonic continuum. In the configuration considered here, distinct transparency dips appear at 850, 865, 870, and 880 nm for detuning values of 0, 26.4, 32.8, and 48.6 meV, respectively. This demonstrates that the toroidal plasmonic continuum can host multiple independently programmable Fano channels within a single nanoantenna platform yielding the possibility of individually addressable switching, or sensing, of molecular resonances within an ensemble.

We emphasize that complete switching of individual quantum modes is only achievable when the QOs are placed within the maximized local-field regions inside or around the TNA. Therefore, non-deterministic QO placement may not produce pronounced spectral dips in the plasmonic continuum. In addition, an excessively large number of QOs or molecular clustering may suppress the visibility of the reported effect by perturbing the local mode structure. Taken together, these results establish the toroidal nanoantenna as a platform for individually addressable quantum-mode switching over a shared plasmonic continuum, with strong modulation depths achievable when deterministic positioning is feasible.

## Conclusion, Discussion and Outlook

In this work, we numerically demonstrated a TNA that emulates a photonic cavity within a specific range of aspect ratios. We developed a comprehensive numerical framework to engineer Fano-resonance phenomena in a hybrid platform composed of a silver plasmonic TNA, a point dipole QE, and one or more narrow Lorentzian-dielectric QOs representing effective molecular-like resonant nanoinclusions. Treating each QO as a discrete narrow-band oscillator evanescently coupled to the TNA, we performed extensive FDTD simulations to:

1. Identify a pronounced Fano transparency dip at $\lambda \approx 850$ nm due to TNA topology, originating from interference between the broadband TNA plasmon continuum and the narrow QO resonance. The efficiency of the modulation depth (over $2840\gamma_0$) yields complete switching and high sensitivity for practical applications.
2. Demonstrate that the Fano-dip transparency reaches a complete transparency (99.9%) for optimal emitter–antenna separations, and TNA provides a large span of distances ($d$ = 3-50 nm) for design freedom in experimental implementations.

3. Introduce multiple QOs within the 3D hotspot region of TNA that broadens the Fano dip when QOs are spectrally resonant and leads to distinct transparency minima when they are spectrally detuned.

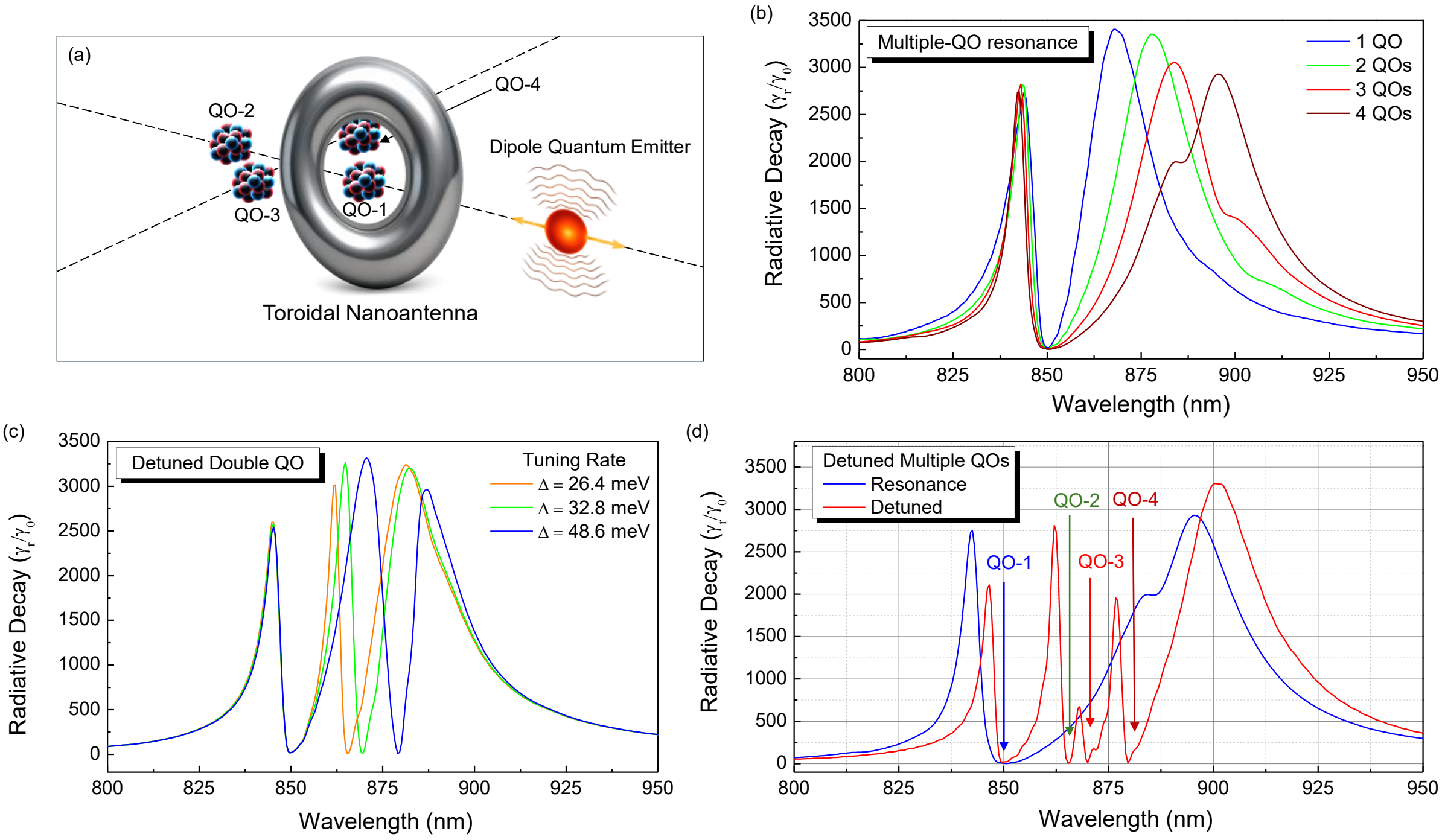

**Figure 4. Individually addressable multi-QO Fano switching in a toroidal plasmonic continuum. (a)** Schematic of the multi-QO configuration, where one QO is positioned at the center of the TNA and additional QOs are placed around the outer hotspot region with fixed lateral separations. The dipolar QE remains at $d = 3$ nm from the antenna. **(b)** Radiative decay spectra for increasing numbers of resonant QOs. When all QOs are tuned to 850 nm, the Fano transparency progressively broadens with increasing QO number, while the transparency wavelength remains nearly unchanged. This demonstrates that multiple narrow resonances can collectively widen the switching bandwidth without substantially shifting the spectral position of the transparency window. **(c)** Radiative decay spectra for the double-QO configuration under spectral detuning. QO-1 is kept resonant at 850 nm, while QO-2 is progressively detuned, causing the emergence and redshift of a second transparency dip. This shows that individual molecular-like QO resonances can imprint distinct and controllable spectral signatures onto the same plasmonic continuum. **(d)** Radiative decay spectra for four spectrally detuned QOs placed within the three-dimensional hotspot region of the TNA. Each QO generates a separate Fano transparency dip at its corresponding resonance wavelength, demonstrating individually addressable switching of multiple quantum modes within a single toroidal plasmonic platform.

Experimental implementation of the presented concept can be realized by coupling a silver TNA to a narrow-line quantum emitter operating in the near-infrared, such as the V1 silicon-vacancy center in 4H-SiC (≈ 861 nm) or a defect center in hBN (550-850 nm band), and probing the predicted Fano-assisted modulation of the emitter's decay rate using a co-localized fluorescent dipole tuned to the same spectral window near 850 nm [63–65]. To that end, we demonstrated an interval that the toroidal geometry remains active for (aspect ratio regime from 0.2 to 0.6), broadband operation with the fabrication of various TNA geometries.

While extending this design into arrays may offer a powerful route toward multifunctional photonic metasurfaces [66], the proposed hybrid system, exhibiting sharp, high-contrast Fano transparency dips in the radiative decay spectrum, is well suited for label-free, real-time biomolecular sensing at the single-molecule level [67]. Moreover, exploiting the presented single and multiple dip concepts within a lithographically compatible toroidal geometry, promises applications in lab-on-chip biosensors with sub-picomolar sensitivity [68]. reconfigurable dual-band filters, slow-light elements for integrated photonics [69,70] and voltage-gated fluorescence switching or dual-colour spacer cavities for quantum-photonic and bio-imaging technologies [71,72]. We further anticipate that suppressing both radiative and non-radiative decay rates traps the energy within the QO instead of re-emitting, thereby possessing possible solutions for programmable emissions in future photonic technologies.

## Methodology

### Analytical framework

To establish a quantitative physical picture of the toroidal nanoantenna (TNA) platform, we approximate the structure by its dominant bright quasinormal mode (QNM), described by annihilation and creation operators $a$and $a^{\dagger}$, with resonance frequency $\omega_c$. As an open and dissipative plasmonic mode, this QNM is characterized by a total linewidth $\kappa = \kappa_{\mathrm{rad}} + \kappa_{\mathrm{nr}}$, where $\kappa_{\mathrm{rad}}$ denotes radiative leakage into free space and $\kappa_{\mathrm{nr}}$ accounts for non-radiative ohmic dissipation in the metal. A two-level QE, with lowering and raising operators $\sigma$and $\sigma^{\dagger}$, transition frequency $\omega_e$, and dipole moment $\boldsymbol{\mu} = \mu\hat{\mathbf{e}}$, is placed at position $\mathbf{r}_e$. The radiative environment is represented by a continuum of free-space modes $\{b_k\}$, while absorptive loss channels are described by a continuum of bath modes $\{c_q\}$. The use of QNMs and their normalization for open nanophotonic resonators follows established treatments for lossy and leaky systems [73,74].

Within the weak-drive, linear-response regime and under the rotating-wave approximation, the Hamiltonian can be written as

$$\frac{H}{\hbar} = \omega_c a^{\dagger} a + \omega_e \sigma^{\dagger}\sigma + (g a \sigma^{\dagger} + \mathrm{h.c.}) + \sum_k \omega_k\, b_k^{\dagger} b_k + \sum_q \omega_q\, c_q^{\dagger} c_q$$

$$+\sum_k(\eta_k a^\dagger b_k + \text{h.c.}) + \sum_q(\eta_q a^\dagger c_q + \text{h.c.}). \tag{2}$$

The QE-mode coupling strength is determined by the local QNM field profile,

$$g = \frac{\mu\cdot\mathbf{E}_c(\mathbf{r}_e)}{\hbar}\sqrt{\frac{\hbar\omega_c}{2\varepsilon_0 U_c}}, U_c \propto V_{\text{eff}}, \tag{3}$$

so that

$$g^2 \propto \frac{\mu^2}{V_{\text{eff}}} \mid \hat{\mathbf{e}}\cdot\mathbf{E}_c(\mathbf{r}_e) \mid^2. \tag{4}$$

Here, $\mathbf{E}_c(\mathbf{r})$is the properly normalized QNM field, $U_c$ is the QNM normalization energy, and $V_{\text{eff}}$ is the effective mode volume at the emitter position [74,75]. This form makes it explicit that strong coupling to the toroidal mode is promoted by both intense local field confinement and small effective modal volume.

The spontaneous emission rate of the emitter in an inhomogeneous electromagnetic environment is governed by the imaginary part of the Green tensor $G_{ij}$, according to

$$\gamma_{\text{tot}}(\omega) = \frac{2\omega^2}{\hbar\varepsilon_0 c^2}\,\mu_i\mu_j \,\text{Im}\, G_{ij}(\mathbf{r}_e,\mathbf{r}_e;\omega). \tag{5}$$

Near a single dominant bright QNM, the Green tensor assumes a Lorentzian spectral form,

$$\text{Im}\, G(\mathbf{r}_e,\mathbf{r}_e;\omega) \simeq \frac{A(\mathbf{r}_e)\left(\frac{\kappa}{2}\right)}{(\omega-\omega_c)^2+(\kappa/2)^2}, \qquad A(\mathbf{r}_e) \propto \frac{\mid \hat{\mathbf{e}}\cdot\mathbf{E}_c(\mathbf{r}_e) \mid^2}{V_{\text{eff}}}. \tag{6}$$

At resonance ( $\omega = \omega_c$ ), this yields the scaling

$$\gamma_{\text{tot}}^{(\text{res})} \propto \frac{\mu^2}{\hbar^2}\,\frac{1}{V_{\text{eff}}}\,\frac{1}{\kappa}, \tag{7}$$

which highlights the central design principle of the TNA platform: maximizing the emission response requires simultaneously reducing the effective mode volume and maintaining a sufficiently narrow bright-mode linewidth.

This interpretation is consistent with the Purcell picture,

$$F_P \equiv \frac{\gamma_{\mathrm{tot}}}{\gamma_0} = \frac{3}{4\pi^2}\left(\frac{\lambda}{n}\right)^3 \frac{Q}{V_{\mathrm{eff}}}, \qquad Q \equiv \frac{\omega_c}{\kappa}, \tag{8}$$

which we use here as an instructive scaling relation. Although this canonical form is exact only in the high-$Q$, Hermitian limit, its QNM generalization remains a useful framework for interpreting emission enhancement in lossy plasmonic resonators.

Because the modal energy can decay through both radiative and ohmic channels, the total spontaneous-emission rate can be partitioned as

$$\gamma_{\mathrm{rad}}^{(\mathrm{res})} = \beta\, \gamma_{\mathrm{tot}}^{(\mathrm{res})}, \qquad \gamma_{\mathrm{nr}}^{(\mathrm{res})} = (1-\beta)\, \gamma_{\mathrm{tot}}^{(\mathrm{res})}, \quad \beta = \frac{\kappa_{\mathrm{rad}}}{\kappa_{\mathrm{rad}} + \kappa_{\mathrm{nr}}}. \tag{9}$$

Accordingly, the radiative Purcell factor becomes

$$F_{\mathrm{rad}} = \beta\, F_P = \beta \frac{3}{4\pi^2}\left(\frac{\lambda}{n}\right)^3 \frac{Q}{V_{\mathrm{eff}}}. \tag{10}$$

This expression makes clear that strong radiative-rate enhancement requires not only subwavelength confinement and favorable modal quality, but also an efficient radiative branching ratio, so that energy coupled into the bright toroidal mode is preferentially emitted rather than dissipated ohmically.

To account for Fano interference, we introduce a narrow discrete resonance with annihilation operator $d$, resonance frequency $\omega_0$, and intrinsic linewidth $\Gamma$, coupled to the bright QNM with interaction strength $J$. The additional Hamiltonian term is

$$\frac{H_{\mathrm{narrow}}}{\hbar} = \omega_0 d^{\dagger} d + (J a^{\dagger} d + \mathrm{h.c.}). \tag{11}$$

Under this single-mode dressed-state picture, the effective cavity susceptibility takes the form

$$\chi_c(\omega) = \frac{1}{\omega - \omega_c + i\kappa/2 - \dfrac{|\,J\,|^2}{\omega - \omega_0 + i\Gamma/2}}, \tag{12}$$

showing that the broad bright toroidal mode interferes with a narrow discrete pathway. This interference gives rise to the characteristic asymmetric Fano line shape and, under appropriate coupling and detuning conditions, to sharp transparency minima in the decay spectrum.

In the present system, the quantum objects (QOs) are modeled as dielectric nano-inclusions with narrow Lorentzian resonances. For inclusion volumes $V_i \ll \lambda^3$and to leading order in the dielectric perturbation $\Delta\varepsilon_i(\omega)$, the Lippmann–Schwinger (Dyson) equation yields a first-order correction to the toroid-only Green tensor [76,77],

$$G(\mathbf{r}, \mathbf{r}'; \omega) \simeq G^{(T)}(\mathbf{r}, \mathbf{r}'; \omega) + k_0^2 \sum_i \Delta\,\varepsilon_i V_i\, G^{(T)}(\mathbf{r}, \mathbf{r}_i; \omega)\, G^{(T)}(\mathbf{r}_i, \mathbf{r}'; \omega), \tag{13}$$

where $G^{(T)}$denotes the Green tensor of the isolated toroidal nanoantenna. We emphasize that Eq. (13) is used only as a weak-perturbation conceptual expression to illustrate how a narrow resonant object can introduce an additional scattering pathway into the toroidal Green tensor. For the Lorentzian parameters used in the FDTD simulations, the resonant dielectric contrast can become large, and therefore the first-order Born approximation is not used as a quantitative model of the QO response. The quantitative decay-rate spectra reported in this work are obtained from full-wave FDTD simulations of the complete dispersive inclusion. Conceptually, the smooth toroidal LDOS can interfere with the narrow Lorentzian resonance introduced by the QO, producing a Fano-type spectral response,

$$\frac{\gamma_{r,\mathrm{wQO}}}{\gamma_0} \propto \frac{(\epsilon+q)^2}{1+\epsilon^2}, \tag{14}$$

with

$$\epsilon = \frac{\omega - \omega_{\mathrm{QO}} - \Delta(\omega)}{\Gamma_{\mathrm{QO}}/2}. \tag{15}$$

Here, $q$ is determined by the complex overlap of the toroidal Green tensor pathways, while $\Delta(\omega)$ represents the real part of the self-energy shift. In the weak-perturbation limit, Eq. (13) illustrates how two or more detuned QOs can introduce additional interference pathways embedded within the same toroidal electromagnetic background. These terms are interpreted as classical or semiclassical interference among several discrete resonant channels, rather than as signatures of many-body quantum correlations.

**Numerical methodology**

Full-wave simulations were performed using a three-dimensional finite-difference time-domain (FDTD) solver (Ansys Lumerical FDTD Solutions) to solve Maxwell's curl equations for the hybrid QE-TNA-QO system. Spatial and temporal discretization were chosen to satisfy the Courant stability criterion, and perfectly matched layer (PML) boundary conditions were employed in all directions to reproduce open-system behavior.

The optical response of the QOs was modeled using a Lorentzian dielectric function of the form

$$\varepsilon(\omega) = \varepsilon_\infty + \frac{f\omega_0^2}{\omega_0^2 - \omega^2 - i\gamma\omega}, \tag{16}$$

where $\varepsilon_\infty$ is the background permittivity, $f$ is the Lorentz oscillator strength, $\omega_0$ is the resonance frequency, and $\gamma$ is the Lorentzian linewidth. Unless otherwise stated, each QO was taken as a spherical dielectric inclusion with radius 20 nm, oscillator strength 0.2, and linewidth on the order of $10^{10}$ rad/s, chosen to represent a narrow molecular-like transition. The Lorentzian dielectric function in Eq. (16) defines the dispersive material response of the QO. For a finite spherical QO, however, the electromagnetic response is more appropriately interpreted through its effective polarizability,

$$\alpha_{\mathrm{QO}}(\omega) = 4\pi\varepsilon_0\varepsilon_b a^3 \frac{\varepsilon(\omega) - \varepsilon_b}{\varepsilon(\omega) + 2\varepsilon_b}, \tag{17}$$

where $a$ is the QO radius and $\varepsilon_b$ is the background permittivity. This form includes the depolarization denominator of a spherical inclusion. Therefore, the first-order Green-tensor expression in Eq. (13) is used only as a weak-perturbation conceptual picture, while the quantitative results are obtained from full-wave FDTD simulations of the complete dispersive QO. The silver TNA, whose geometry is defined by a major radius $R$ and a cross-sectional radius $\alpha$, was described using the experimental optical constants reported by Johnson and Christy [78]. Johnson and Christy optical constants were used as an experimentally grounded local-response model for silver, since previous experiment-simulation studies on Ag nanostructures have shown that these data can reproduce measured plasmonic spectra reasonably well, although nonlocal, surface-scattering, and quantum-corrected effects may still modify the absolute enhancement values at the smallest dimensions [79,80]. Because the smallest emitter-antenna separation considered here is 3 nm and no direct metal-metal subnanometer junction is present, electron tunneling is not explicitly included [81].

The QE was modeled as an electric point dipole emitter operating near 850 nm for the aspect-ratio configuration $\alpha/R = 0.2$. For other geometries, the emitter wavelength was adjusted to probe the corresponding resonance condition of the TNA. To ensure numerical convergence in the strongly confined near-field region, a fine nonuniform mesh was employed, with local cell sizes ranging from approximately $\lambda/2600$ to $\lambda/600$, depending on the emitter-antenna separation and the associated near-field gradients.

The free-space spontaneous emission rate is given by

$$\gamma_0 = \frac{\omega^3 \mid \mathbf{p} \mid^2}{3\pi\varepsilon_0\hbar c^3}, \tag{18}$$

where $\omega$ is the angular frequency and $\mathbf{p}$ is the dipole moment. In the presence of the TNA-QO environment, the radiative decay rate was extracted from far-field power monitors according to

$$\gamma_{\text{rad}} = \frac{P_{\text{rad}}}{P_0} \gamma_0, \tag{19}$$

where $P_{\text{rad}}$ is the radiated power collected in the far field and $P_0$is the power emitted by the same dipole in free space. The total decay rate was obtained from the total power emitted into the simulation domain,

$$\gamma_{\text{tot}} = \frac{P_{\text{tot}}}{P_0} \gamma_0, \tag{20}$$

with $P_{\text{tot}}$ evaluated from the total electromagnetic power flux through a closed surface surrounding the dipole and the hybrid nanostructure. The non-radiative decay rate then follows directly as

$$\gamma_{\text{nr}} = \gamma_{\text{tot}} - \gamma_{\text{rad}}. \tag{21}$$

Specifically, the total emitted power is calculated from the surface integral of the Poynting vector over a closed near-field surface $S$,

$$P_{\text{tot}} = \oint_S \mathbf{S} \cdot d\mathbf{A}, \tag{22}$$

while the radiated power is obtained from an analogous integration over a far-field monitor surface $S_{\text{far}}$,

$$P_{\text{rad}} = \oint_{S_{\text{far}}} \mathbf{S} \cdot d\mathbf{A}. \tag{23}$$

This separation allows direct discrimination between power emitted to the far field and power dissipated non-radiatively through absorption or near-field energy transfer processes.

To parameterize the distance dependence of the QE-TNA interaction, the decay-rate variation with emitter–antenna separation $x$was fitted using an empirical exponential form,

$$y_i(x) = \exp\left(a + bx + cx^2\right), \tag{24}$$

where the fitting coefficients were obtained separately for the radiative and non-radiative channels. For the radiative decay rate, the coefficients are $a = 8.40839$, $b = -0.16056$, and $c = 0.00162$, whereas for the non-radiative decay rate they are $a = 7.52683$, $b = -0.20485$, and $c = 0.00243$. These fits are used only to parameterize the simulated distance dependence and do not enter the electromagnetic solver itself.

## Acknowledgments

E.O.P. acknowledges financial support from the Scientific and Technological Research Council of Türkiye (TÜBİTAK) through Grant No. 222N308 within the CHIST-ERA program and BILKENT University- TUBITAK BILGEM Consultancy Call for Research (BIL2). Additional support was provided by the Turkish Academy of Sciences Outstanding Young Scientists Awards (GEBİP) 2025.

## Author Contributions

A.G. performed the numerical simulations, carried out the calculations, and extracted the raw data underlying the reported results. A.G. also contributed to the preparation of the manuscript. E.O.P. conceived the research idea, supervised the project, and led the preparation of the manuscript, including the development of the final figures and the main text. All authors discussed the results and contributed to the final version of the manuscript.

## Competing interests

The authors declare no competing interests.

## Data availability

All data necessary to evaluate the results presented in this work are included within the paper. Additional datasets supporting the findings of this study are available from the corresponding author upon reasonable request.

## Refererences

[1] Y. Wy, H. Jung, J. W. Hong, and S. W. Han, Exploiting Plasmonic Hot Spots in Au-Based Nanostructures for Sensing and Photocatalysis, Acc. Chem. Res. 55, 831 (2022).

[2] S. A. Maier, M. L. Brongersma, P. G. Kik, S. Meltzer, A. A. G. Requicha, and H. A. Atwater, Plasmonics—A Route to Nanoscale Optical Devices, Advanced Materials 13, 1501 (2001).

[3] S. A. Maier and others, *Plasmonics: Fundamentals and Applications*, Vol. 1 (Springer, 2007).

[4] K. Kneipp, Y. Wang, H. Kneipp, L. T. Perelman, I. Itzkan, R. R. Dasari, and M. S. Feld, Single Molecule Detection Using Surface-Enhanced Raman Scattering (SERS), Phys. Rev. Lett. 78, 1667 (1997).

[5] E. Ozbay, Plasmonics: Merging photonics and electronics at nanoscale dimensions, Science (1979). 311, 189 (2006).

[6] A. Gulucu and E. O. Polat, Optically Switchable Fluorescence Enhancement at Critical Interparticle Distances, Adv. Theory Simul. 8, e01134 (2025).

[7] R. Sharma, N. K. Pathak, and R. P. Sharma, Computational Study of Plasmon Interaction in Organic Media: a Comparison Between Analytical and Numerical Model for Dimer, Plasmonics 13, 1775 (2018).

[8] Y. F. Xiao, Y. C. Liu, B. B. Li, Y. L. Chen, Y. Li, and Q. Gong, Strongly enhanced light-matter interaction in a hybrid photonic-plasmonic resonator, Phys. Rev. A 85, 031805 (2012).

[9] J. Cao, T. Sun, and K. T. V. Grattan, Gold nanorod-based localized surface plasmon resonance biosensors: A review, Sens. Actuators B Chem. 195, 332 (2014).

[10] T. Kaelberer, V. A. Fedotov, N. Papasimakis, D. P. Tsai, and N. I. Zheludev, Toroidal Dipolar Response in a Metamaterial, Science (1979). 330, 1510 (2010).

[11] N. Papasimakis, V. A. Fedotov, V. Savinov, T. A. Raybould, and N. I. Zheludev, Electromagnetic toroidal excitations in matter and free space, Nat. Mater. 15, 263 (2016).

[12] A. Mary, A. Dereux, and T. L. Ferrell, Localized surface plasmons on a torus in the nonretarded approximation, Phys. Rev. B Condens. Matter Mater. Phys. 72, 155426 (2005).

[13] T. Warnakula, S. D. Gunapala, M. I. Stockman, and M. Premaratne, Broken poloidal symmetry and plasmonic eigenmodes on a torus, Phys. Rev. B 101, 115426 (2020).

[14] T. V. Teperik and A. Degiron, Numerical analysis of an optical toroidal antenna coupled to a dipolar emitter, Phys. Rev. B Condens. Matter Mater. Phys. 83, 245408 (2011).

[15] C. M. Dutta, T. A. Ali, D. W. Brandl, T. H. Park, and P. Nordlander, Plasmonic properties of a metallic torus, Journal of Chemical Physics 129, (2008).

[16] J. Aizpurua, P. Hanarp, D. S. Sutherland, M. Käll, G. W. Bryant, and F. J. García de Abajo, Optical Properties of Gold Nanorings, Phys. Rev. Lett. 90, 4 (2003).

[17] H. E. M. Purcell, Spontaneous Emission Probabilities at Radio Frequencies, 839 (1995).

[18] K. V. Garapati, M. Salhi, S. Kouchekian, G. Siopsis, and A. Passian, Poloidal and toroidal plasmons and fields of multilayer nanorings, Phys. Rev. B 95, 165422 (2017).

[19] M. F. Limonov, M. V. Rybin, A. N. Poddubny, and Y. S. Kivshar, Fano resonances in photonics, Nat. Photonics 11, 543 (2017).

[20] U. Fano, Effects of Configuration Interaction on Intensities and Phase Shifts, Physical Review 124, 1866 (1961).

[21] M. Pelton, S. D. Storm, and H. Leng, Strong coupling of emitters to single plasmonic nanoparticles: exciton-induced transparency and Rabi splitting, Nanoscale 11, 14540 (2019).

[22] N. Liu, L. Langguth, T. Weiss, J. Kästel, M. Fleischhauer, T. Pfau, and H. Giessen, Plasmonic analogue of electromagnetically induced transparency at the Drude damping limit, Nat. Mater. 8, 758 (2009).

[23] H. Leng, B. Szychowski, M. C. Daniel, and M. Pelton, Strong coupling and induced transparency at room temperature with single quantum dots and gap plasmons, Nature Communications 2018 9:1 9, 1 (2018).

[24] S. Zhang, D. A. Genov, Y. Wang, M. Liu, and X. Zhang, Plasmon-induced transparency in metamaterials, Phys. Rev. Lett. 101, 047401 (2008).

[25] Y. Zheng et al., Fano Resonance in Single-Molecule Junctions, Angew. Chem. Int. Ed. 61, e202210097 (2022).

[26] A. E. Miroshnichenko, S. Flach, and Y. S. Kivshar, Fano resonances in nanoscale structures, Rev. Mod. Phys. 82, 2257 (2010).

[27] B. Luk'Yanchuk, N. I. Zheludev, S. A. Maier, N. J. Halas, P. Nordlander, H. Giessen, and C. T. Chong, The Fano resonance in plasmonic nanostructures and metamaterials, Nat. Mater. 9, 707 (2010).

[28] Q. Shi, Z. Yu, Y. Liu, H. Gong, H. Yin, W. Zhang, J. Liu, and Y. Peng, Plasmonics properties of nano-torus: An FEM method, Opt. Commun. 285, 4542 (2012).

[29] M. Pelton, S. K. Gray, and X. Wu, Quantum-dot-induced transparency in a nanoscale plasmonic resonator, Optics Express, Vol. 18, Issue 23, Pp. 23633-23645 18, 23633 (2010).

[30] J. Yan, C. Ma, P. Liu, C. Wang, and G. Yang, Generating scattering dark states through the Fano interference between excitons and an individual silicon nanogroove, Light: Science & Applications 2017 6:1 6, e16197 (2016).

[31] R. A. Shah, N. F. Scherer, M. Pelton, and S. K. Gray, Ultrafast reversal of a Fano resonance in a plasmon-exciton system, Phys. Rev. B 88, 075411 (2013).

[32] M. Dowran, U. Kilic, S. Lamichhane, A. Erickson, J. Barker, M. Schubert, S. H. Liou, C. Argyropoulos, and A. Laraoui, Plasmonic Nanocavity to Boost Single Photon Emission From Defects in Thin Hexagonal Boron Nitride, Laser Photon. Rev. 19, 2400705 (2025).

[33] X. Xu et al., Greatly Enhanced Emission from Spin Defects in Hexagonal Boron Nitride Enabled by a Low-Loss Plasmonic Nanocavity, Nano Lett. 23, 25 (2022).

[34] D. E. Westmoreland, K. P. McClelland, K. A. Perez, J. C. Schwabacher, Z. Zhang, and E. A. Weiss, Properties of quantum dots coupled to plasmons and optical cavities, Journal of Chemical Physics 151, 210901 (2019).

[35] S. Hu, J. Huang, R. Arul, A. Sánchez-Iglesias, Y. Xiong, L. M. Liz-Marzán, and J. J. Baumberg, Robust consistent single quantum dot strong coupling in plasmonic nanocavities, Nature Communications 2024 15:1 15, 6835 (2024).

[36] A. Hassanfiroozi, P. S. Huang, S. H. Huang, K. I. Lin, Y. T. Lin, C. F. Chien, Y. Shi, W. J. Lee, and P. C. Wu, A Toroidal-Fano-Resonant Metasurface with Optimal Cross-Polarization Efficiency and Switchable Nonlinearity in the Near-Infrared, Adv. Opt. Mater. 9, 2101007 (2021).

[37] Q. Mi, T. Sang, Y. Pei, C. Yang, S. Li, Y. Wang, and B. Ma, High-quality-factor dual-band Fano resonances induced by dual bound states in the continuum using a planar nanohole slab, Nanoscale Research Letters 2021 16:1 16, 150 (2021).

[38] N. Papasimakis, V. A. Fedotov, N. I. Zheludev, and S. L. Prosvirnin, Metamaterial Analog of Electromagnetically Induced Transparency, Phys. Rev. Lett. 101, 253903 (2008).

[39] E. O. Polat, Z. Artvin, Y. Şaki, A. Bek, and R. Sahin, Continuous and reversible electrical-tuning of fluorescent decay rate via Fano resonance, Nanotechnology 37, 165701 (2026).

[40] P. C. Wu et al., Optical Anapole Metamaterial, ACS Nano 12, 1920 (2018).

[41] M. J. Urban, P. K. Dutta, P. Wang, X. Duan, X. Shen, B. Ding, Y. Ke, and N. Liu, Plasmonic Toroidal Metamolecules Assembled by DNA Origami, J. Am. Chem. Soc. 138, 5495 (2016).

[42] W. Pfaff, A. Vos, and R. Hanson, Top-down fabrication of plasmonic nanostructures for deterministic coupling to single quantum emitters, J. Appl. Phys. 113, 24310 (2013).

[43] Z. Zhao, X. Chen, J. Zuo, A. Basiri, S. Choi, Y. Yao, Y. Liu, and C. Wang, Deterministic assembly of single emitters in sub-5 nanometer optical cavity formed by gold nanorod dimers on three-dimensional DNA origami, Nano Research 2021 15:2 15, 1327 (2021).

[44] T. T. Tran, D. Wang, Z. Q. Xu, A. Yang, M. Toth, T. W. Odom, and I. Aharonovich, Deterministic Coupling of Quantum Emitters in 2D Materials to Plasmonic Nanocavity Arrays, Nano Lett. 17, 2634 (2017).

[45] D. A. B. Miller, D. S. Chemla, T. C. Damen, A. C. Gossard, W. Wiegmann, T. H. Wood, and C. A. Burrus, Band-Edge Electroabsorption in Quantum Well Structures: The Quantum-Confined Stark Effect, Phys. Rev. Lett. 53, 2173 (1984).

[46] A. M. Armani, R. P. Kulkarni, S. E. Fraser, R. C. Flagan, and K. J. Vahala, Label-free, single-molecule detection with optical microcavities, Science (1979). 317, 783 (2007).

[47] F. A. S. Barbosa, A. S. Coelho, K. N. Cassemiro, P. Nussenzveig, C. Fabre, M. Martinelli, and A. S. Villar, Beyond Spectral Homodyne Detection: Complete Quantum Measurement of Spectral Modes of Light, Phys. Rev. Lett. 111, 200402 (2013).

[48] Y. Miao, R. C. Boutelle, A. Blake, V. Chandrasekaran, C. J. Sheehan, J. Hollingsworth, D. Neuhauser, and S. Weiss, Super-resolution Imaging of Plasmonic Near-Fields: Overcoming Emitter Mislocalizations, J. Phys. Chem. Lett. 13, 4520 (2022).

[49] C. Lee, B. Lawrie, R. Pooser, K. G. Lee, C. Rockstuhl, and M. Tame, Quantum Plasmonic Sensors, Chem. Rev. 121, 4743 (2021).

[50] W. Bogaerts, D. Pérez, J. Capmany, D. A. B. Miller, J. Poon, D. Englund, F. Morichetti, and A. Melloni, Programmable photonic circuits, Nature 2020 586:7828 586, 207 (2020).

[51] X. Duan, S. Kamin, and N. Liu, Dynamic plasmonic colour display, Nature Communications 2017 8:1 8, 1 (2017).

[52] L. Novotny and B. Hecht, Principles of Nano-Optics, Principles of Nano-Optics 9781107005464, 1 (2012).

[53] D. Franklin, R. Frank, S. T. Wu, and D. Chanda, Actively addressed single pixel full-colour plasmonic display, Nature Communications 2017 8:1 8, 15209 (2017).

[54] L. A. Mayoral Astorga et al., Electrically tunable plasmonic metasurface as a matrix of nanoantennas, Nanophotonics 13, 901 (2024).

[55] R. Carminati, J. J. Greffet, C. Henkel, and J. M. Vigoureux, Radiative and non-radiative decay of a single molecule close to a metallic nanoparticle, Opt. Commun. 261, 368 (2006).

[56] P. Anger, P. Bharadwaj, and L. Novotny, Enhancement and Quenching of Single-Molecule Fluorescence Near Metallic Nanostructures, Phys. Rev. Lett. 96, 113002 (2006).

[57] E. Dulkeith, A. C. Morteani, T. Niedereichholz, T. A. Klar, J. Feldmann, S. A. Levi, F. C. J. M. van Veggel, D. N. Reinhoudt, M. Möller, and D. I. Gittins, Fluorescence Quenching of Dye Molecules near Gold Nanoparticles: Radiative and Nonradiative Effects, Phys. Rev. Lett. 89, 203002 (2002).

[58] G. P. Acuna et al., Distance Dependence of Single-Fluorophore Quenching by Gold Nanoparticles Studied on DNA Origami, ACS Nano 6, 3189 (2012).

[59] E. Wientjes, J. Renger, A. G. Curto, R. Cogdell, and N. F. Van Hulst, Nanoantenna enhanced emission of light-harvesting complex 2: the role of resonance, polarization, and radiative and non-radiative rates, Physical Chemistry Chemical Physics 16, 24739 (2014).

[60] E. Wientjes, J. Renger, A. G. Curto, R. Cogdell, and N. F. Van Hulst, Strong antenna-enhanced fluorescence of a single light-harvesting complex shows photon antibunching, Nature Communications 2014 5:1 5, 4236 (2014).

[61] G. W. Wen, J. Y. Lin, H. X. Jiang, and Z. Chen, Quantum-confined Stark effects in semiconductor quantum dots, Phys. Rev. B 52, 5913 (1995).

[62] I. Niehues, E. D. S. Nysten, R. Schmidt, M. Weiß, and D. Wigger, Excitons in quantum technologies: The role of strain engineering, MRS Bulletin 2024 49:9 49, 958 (2024).

[63] H. Bahmani Jalali, L. De Trizio, L. Manna, and F. Di Stasio, Indium arsenide quantum dots: an alternative to lead-based infrared emitting nanomaterials, Chem. Soc. Rev. 51, 9861 (2022).

[64] S. Castelletto, F. A. Inam, S. I. Sato, and A. Boretti, Hexagonal boron nitride: a review of the emerging material platform for single-photon sources and the spin–photon interface, Beilstein Journal of Nanotechnology 11:61 11, 740 (2020).

[65] D. Liu, F. Kaiser, V. Bushmakin, E. Hesselmeier, T. Steidl, T. Ohshima, N. T. Son, J. Ul-Hassan, Ö. O. Soykal, and J. Wrachtrup, The silicon vacancy centers in SiC: determination of intrinsic spin dynamics for integrated quantum photonics, Npj Quantum Information 2024 10:1 10, 72 (2024).

[66] M. Liu et al., Multifunctional metasurfaces enabled by simultaneous and independent control of phase and amplitude for orthogonal polarization states, Light: Science & Applications 2021 10:1 10, 1 (2021).

[67] T. Roesel, A. Dahlin, M. Piliarik, L. W. Fitzpatrick, and B. Špačková, Label-free single-molecule optical detection, Npj Biosensing 2025 2:1 2, 1 (2025).

[68] S. W. Chong, Y. Shen, S. Palomba, and D. Vigolo, Nanofluidic Lab-On-A-Chip Systems for Biosensing in Healthcare, Small 21, 2407478 (2025).

[69] N. A. Salama, S. M. Alexeree, S. S. A. Obayya, and M. A. Swillam, Silicon-based double fano resonances photonic integrated gas sensor, Sci. Rep. 14, (2024).

[70] H. J. Chen, Multiple-Fano-resonance-induced fast and slow light in the hybrid nanomechanical-resonator system, Phys. Rev. A (Coll. Park). 104, 013708 (2021).

[71] Y. Huo, T. Jia, T. Ning, C. Tan, S. Jiang, C. Yang, Y. Jiao, and B. Man, A low lasing threshold and widely tunable spaser based on two dark surface plasmons, Sci. Rep. 7, 1 (2017).

[72] N. K. Emani, T. F. Chung, A. V. Kildishev, V. M. Shalaev, Y. P. Chen, and A. Boltasseva, Electrical modulation of fano resonance in plasmonic nanostructures using graphene, Nano Lett. 14, 78 (2014).

[73] C. Sauvan, J. P. Hugonin, I. S. Maksymov, and P. Lalanne, Theory of the Spontaneous Optical Emission of Nanosize Photonic and Plasmon Resonators, Phys. Rev. Lett. 110, 237401 (2013).

[74] P. T. Kristensen, R. C. Ge, and S. Hughes, Normalization of quasinormal modes in leaky optical cavities and plasmonic resonators, Phys. Rev. A (Coll. Park). 92, 053810 (2015).

[75] E. A. Muljarov and W. Langbein, Exact mode volume and Purcell factor of open optical systems, Phys. Rev. B 94, 235438 (2016).

[76] W. Cho. Chew, Waves and fields in inhomogeneous media, 608 (1999).

[77] O. J. F. Martin, C. Girard, and A. Dereux, Generalized Field Propagator for Electromagnetic Scattering and Light Confinement, Phys. Rev. Lett. 74, 526 (1995).

[78] P. B. Johnson and R. W. Christy, Optical Constants of the Noble Metals, Phys. Rev. B 6, 4370 (1972).

[79] J. Heintz, F. Legittimo, and S. Bidault, Dimers of Plasmonic Nanocubes to Reach Single-Molecule Strong Coupling with High Emission Yields, J. Phys. Chem. Lett. 13, 11996 (2022).

[80] M. D. Wissert, C. Moosmann, K. S. Ilin, M. Siegel, U. Lemmer, and H.-J. Eisler, Gold nanoantenna resonance diagnostics via transversal particle plasmon luminescence, Optics Express, Vol. 19, Issue 4, Pp. 3686-3693 19, 3686 (2011).

[81] K. J. Savage, M. M. Hawkeye, R. Esteban, A. G. Borisov, J. Aizpurua, and J. J. Baumberg, Revealing the quantum regime in tunnelling plasmonics, Nature 2012 491:7425 491, 574 (2012).